\documentclass[journal,twoside,web]{IEEEtran}

\usepackage{cite}
\usepackage{amsmath}
\usepackage[hyphens]{url}
\usepackage{hyperref}
\usepackage{array}
\usepackage{multirow}
\usepackage{graphicx}
\usepackage{gensymb}
\usepackage{float}
\usepackage[super]{nth}
\usepackage{mathtools}
\usepackage{amsfonts}
\usepackage{amssymb}
\usepackage{tikz}

\definecolor{subsectioncolor}{rgb}{0,0,0}

\usepackage{xcolor}

\newcommand\figref[1]{\textcolor{subsectioncolor}{Fig. \ref{#1}}}
\newcommand\figrefmult[2]{\textcolor{subsectioncolor}{Fig. \ref{#1}#2}}
\newcommand\tableref[1]{\textcolor{subsectioncolor}{Table \ref{#1}}}
\newcommand\highlightcolor[1]{\textcolor{subsectioncolor}{#1}}

\makeatletter
\newcommand*{\rom}[1]{\expandafter\@slowromancap\romannumeral #1@}
\makeatother

\newcommand{\mycbox}[1]{\tikz{\path[draw=black,fill=#1] (0,0) rectangle (0.2cm,0.2cm);}}
\newcommand{\mysolidline}[1]{\raisebox{2pt}{\tikz{\draw[-,#1,solid,line width = 1pt](0,0) -- (5mm,0);}}}
\newcommand{\myfilledcircle}[1]{\raisebox{-0.5pt}{\tikz{\node[draw,scale=0.75,circle,fill=#1](){};}}}

\definecolor{shadedFetus}{RGB}{88,104,126}
\definecolor{shadedAmniotic}{RGB}{153,165,105}
\definecolor{shadedAbdomen}{RGB}{150,214,152}
\definecolor{fetus}{RGB}{71,79,255}
\definecolor{vernix}{RGB}{255,255,0}
\definecolor{amniotic}{RGB}{255,79,71}
\definecolor{abdomen}{RGB}{71,255,79}
\definecolor{homogeneous}{RGB}{25.5000,25.5000 ,25.5000}
\definecolor{model1}{RGB}{102.0000,102.0000,102.0000}
\definecolor{model2}{RGB}{0,113.99,188.955}
\definecolor{model3}{RGB}{216.75,82.875,24.99}
\definecolor{model4}{RGB}{255,0,0}
\definecolor{model5}{RGB}{125.97,46.92,141.78}
\definecolor{model6}{RGB}{0,170,0}
\definecolor{model7}{RGB}{76.755,189.975,237.915}
\definecolor{model8}{RGB}{40,40,200}
\definecolor{sensor1}{RGB}{0,0,0}
\definecolor{sensor2}{RGB}{0,0,255}
\definecolor{sensor3}{RGB}{255,102,0}
\definecolor{sensor4}{RGB}{0,255,255}
\definecolor{sensor5}{RGB}{255,0,0}
\definecolor{sensor6}{RGB}{255,0,255}
\definecolor{sensor7}{RGB}{255,255,0}
\definecolor{sensor8}{RGB}{255,255,255}
\definecolor{qrszone}{RGB}{204,204,204}
\definecolor{tzone}{RGB}{229.5,229.5,229.5}

\newcolumntype{C}[1]{>{\centering\let\newline\\\arraybackslash}m{#1}}

\DeclarePairedDelimiter\norm{\lVert}{\rVert}

\begin{document}
\bstctlcite{IEEEtran:bstcontrol}
\title{The effects of asymmetric volume conductor modeling on non-invasive fetal ECG extraction} 
\author{Emerson~Keenan\textsuperscript{\href{https://orcid.org/0000-0003-1966-2293}{\includegraphics[width=8pt]{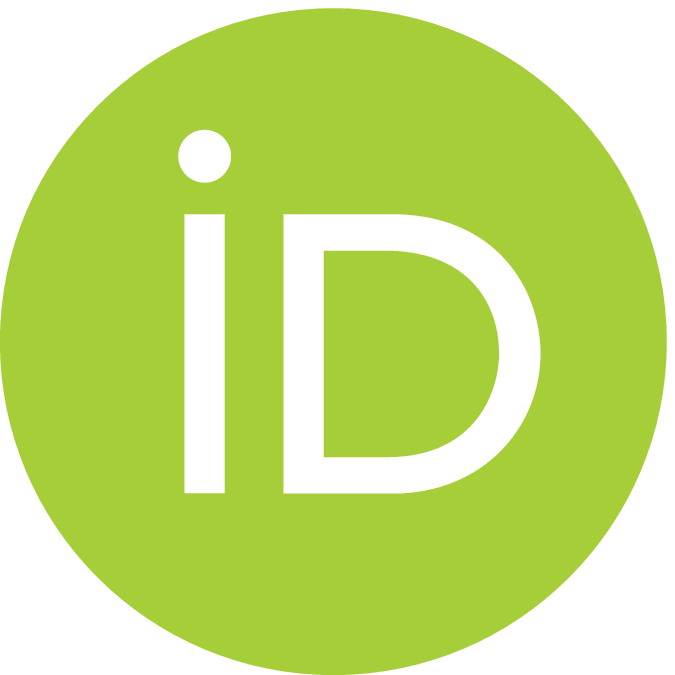}}},~\IEEEmembership{Student Member,~IEEE}, Chandan~Kumar~Karmakar\textsuperscript{\href{https://orcid.org/0000-0003-1814-0856}{\includegraphics[width=8pt]{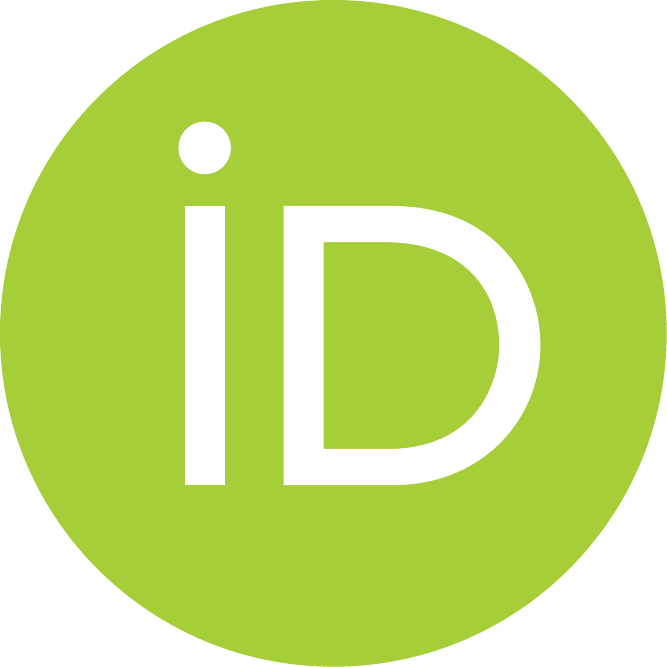}}},~\IEEEmembership{Member,~IEEE}, and Marimuthu~Palaniswami\textsuperscript{\href{https://orcid.org/0000-0002-3635-4252}{\includegraphics[width=8pt]{vector_iD_icon}}},~\IEEEmembership{Fellow,~IEEE} \thanks{Emerson Keenan, Chandan Kumar Karmakar and Marimuthu Palaniswami are with the Department of Electrical and Electronic Engineering, The University of Melbourne, Melbourne, VIC 3010, Australia.} \thanks{Chandan Kumar Karmakar is also with the School of Information Technology, Deakin University, Burwood, VIC 3125, Australia. Correspondence email: \href{mailto:karmakar@deakin.edu.au}{karmakar@deakin.edu.au}}}

\maketitle

\setcounter{footnote}{-1}

\begin{abstract}
	\textit{Objective:} Non-invasive fetal electrocardiography (NI-FECG) shows promise for capturing novel physiological information that may indicate signs of fetal distress. However, significant deterioration in NI-FECG signal quality occurs during the presence of a highly non-conductive layer known as vernix caseosa which forms on the fetal body surface beginning in approximately the \nth{28} week of gestation. This work investigates asymmetric modeling of vernix caseosa and other maternal-fetal tissues in accordance with clinical observations and assesses their impacts for NI-FECG signal processing. \textit{Methods:} We develop a process for simulating dynamic maternal-fetal abdominal ECG mixtures using a synthetic cardiac source model embedded in a finite element volume conductor. Using this process, changes in NI-FECG signal morphology are assessed in an extensive set of finite element models including spatially variable distributions of vernix caseosa. \textit{Results:} Our simulations show that volume conductor asymmetry can result in over 70\% error in the observed T/QRS ratio and significant changes to signal morphology compared to a homogeneous volume conductor model. \textit{Conclusion:} Volume conductor effects must be considered when analyzing T/QRS ratios obtained via NI-FECG and should be considered in future algorithm benchmarks using simulated data. \textit{Significance:} This work shows that without knowledge of the influence of volume conductor effects, clinical evaluation of the T/QRS ratio derived via NI-FECG should be avoided.
\end{abstract}
\vspace*{0.2cm}
\begin{IEEEkeywords}
	Finite element method, non-invasive fetal electrocardiography (NI-FECG), T/QRS ratio, vernix caseosa, volume conductor model.
\end{IEEEkeywords}
\vspace*{0.4cm}
\section{Introduction}
\IEEEPARstart{N}{on-invasive} fetal electrocardiography (NI-FECG) has shown promise in recent years with several NI-FECG systems becoming commercially available for monitoring fetal heart rate (FHR) in a clinical environment \cite{Sameni_ReviewFetalECG_2010,CohenWayneR._Accuracyreliabilityfetal_2012,Clifford_NoninvasivefetalECG_2014}. NI-FECG demonstrates several advantages over ultrasound based cardiotocography (CTG) in that it is a passive modality suitable for long term use and may provide novel physiological information in addition to FHR \cite{Symonds_Fetalelectrocardiography_2001,Clifford_Clinicallyaccuratefetal_2011,Behar_EvaluationfetalQT_2016}. As the ECG waveform aims to represent the magnitude and temporal characteristics of electrical currents within the cardiac muscle, morphological analysis of NI-FECG recordings has been proposed for identifying signs of fetal distress such as hypoxia and intrauterine growth restriction (IUGR)
\cite{Behar_practicalguidenoninvasive_2016,Oudijk_effectsintrapartumhypoxia_2004,Fuchs_ValuesQRSratio_2016}.

\begin{figure}
	\centering
	\includegraphics[width=0.5\textwidth]{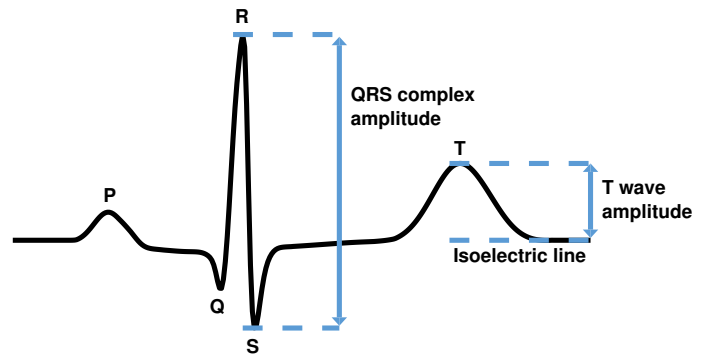}
	\caption{An exemplary ECG waveform segment indicating the P wave, QRS complex and T wave. A proposed morphological feature for identifying fetal distress is the ratio of T wave amplitude to QRS complex amplitude (T/QRS ratio).}
	\label{tqrs}
	\vspace*{-0.2cm}
\end{figure}

The process of morphological analysis can be divided into two distinct phases: 1) Extraction - the process of extracting a fetal ECG waveform from noisy single or multi-channel recordings and 2) Analysis - the process of analyzing features of the extracted ECG waveform to assess physiological state. These features can be separated into the characteristics of individual physiological events (e.g. QRS complex amplitude and location) and the ratios or intervals between events. \mbox{\figref{tqrs}} shows an exemplary ECG waveform segment indicating one such feature, the ratio of T wave amplitude to QRS complex amplitude (T/QRS ratio) which has been proposed for identifying fetal distress \cite{Behar_practicalguidenoninvasive_2016,Fuchs_ValuesQRSratio_2016}. The accuracy of morphological analysis therefore depends on the accuracy of the extraction process.

However, extracting an accurate fetal ECG from a set of abdominal recordings is challenging due to interference from the maternal ECG, muscle activity, fetal brain activity and sensor noise \cite{Stinstra_InfluenceFetoabdominalTissues_2002,Andreotti_RobustfetalECG_2014}. In addition to these disturbances, a highly non-conductive layer known as \textit{vernix caseosa} forms on the fetal body surface beginning in approximately the \nth{28} week of gestation causing significant attenuation of fetal cardiac signals \cite{Stinstra_reliabilityfetalmagnetocardiogram_2001,Oostendorp_effectchangesconductive_1989}. The influence of vernix caseosa presents an important area of research for NI-FECG extraction as many signal processing approaches used in this domain (e.g. adaptive filtering \cite{Widrow_Adaptivenoisecancelling_1975}, template subtraction \cite{Martens_robustfetalECG_2007}) consider the maternal abdomen as a homogeneous volume conductor. This assumes that recordings at the skin surface contain a superimposed projection of underlying electrical sources with attenuation as a function of distance from source to sensor position.

By comparison, attenuation in an inhomogeneous volume conductor depends on the specific position and orientation of each source in the 3D domain as well as its distance to the sensor position. As the cardiac events represented by the P wave, QRS complex and T wave of the ECG waveform are oriented differently in 3D space \cite{Nousiainen_Comparativestudynormal_1986}, it is important to determine if the presence of vernix caseosa impacts NI-FECG extraction accuracy through the use of appropriate volume conductor models.

Prior work by Stinstra et al. \cite{Stinstra_InfluenceFetoabdominalTissues_2002} and Oostendorp et al. \cite{Oostendorp_effectchangesconductive_1989} investigated the effects of vernix caseosa on abdominal potentials by modeling it as a uniform layer covering the fetal body surface with holes near the mouth and umbilicus. Oostendorp et al. \cite{Oostendorp_effectchangesconductive_1989} concluded that while this model approximated some instances of clinical NI-FECG data, it did not account for all observations due to possible variations in vernix caseosa distribution. These works focused on relatively symmetric distributions of vernix caseosa and did not investigate observations presented by Akiba \cite{Akiba_Studiesbiologicalactions_1955} on a cohort of newborns ($n=623$) from 28 weeks gestational age (GA) onwards which found that while 16\% of the cohort had vernix caseosa covering the entire body, the remaining group had either no vernix caseosa or distributions favouring the following regions in order of decreasing prevalence: inguinal region, axillary fossa, back, buttocks, hips, thighs and neck. Preferential surface distributions of vernix caseosa have also been reported by Archana \cite{Archana_ClinicalStudySurface_2008} where vernix caseosa in the cohort ($n=100$) was found in 78\% of cases on the back, 61\% in the inguinal region and 16\% on the chest, and by Visscher et al. \cite{Visscher_VernixCaseosaNeonatal_2005} where vernix caseosa in the cohort ($n=430$) was significantly higher on the back than the chest. These results indicate that modeling vernix caseosa as a uniform layer does not represent a significant proportion of cases and a non-uniform, asymmetric model is required to fully understand its impacts on NI-FECG extraction. The contributions of this work are thus as follows:
\begin{itemize}
	\item Link clinical observations of the asymmetric distribution of vernix caseosa to their relevance in the study of NI-FECG signal processing
	\item Develop a novel process to simulate dynamic maternal-fetal ECG mixtures in an asymmetric volume conductor model including spatially variable distributions of vernix caseosa
	\item Characterise the effects of asymmetric volume conductor models on NI-FECG extraction in terms of changes to signal morphology compared to a homogeneous volume conductor model
	\item Demonstrate that NI-FECG algorithm benchmarks using a homogeneous volume conductor model do not provide representative accuracy of T/QRS ratio extraction
	\item Provide recommendations for the utility of the T/QRS ratio derived via NI-FECG as a clinical tool
\end{itemize}

\begin{figure*}
	\centering
	\includegraphics[width=0.88\textwidth]{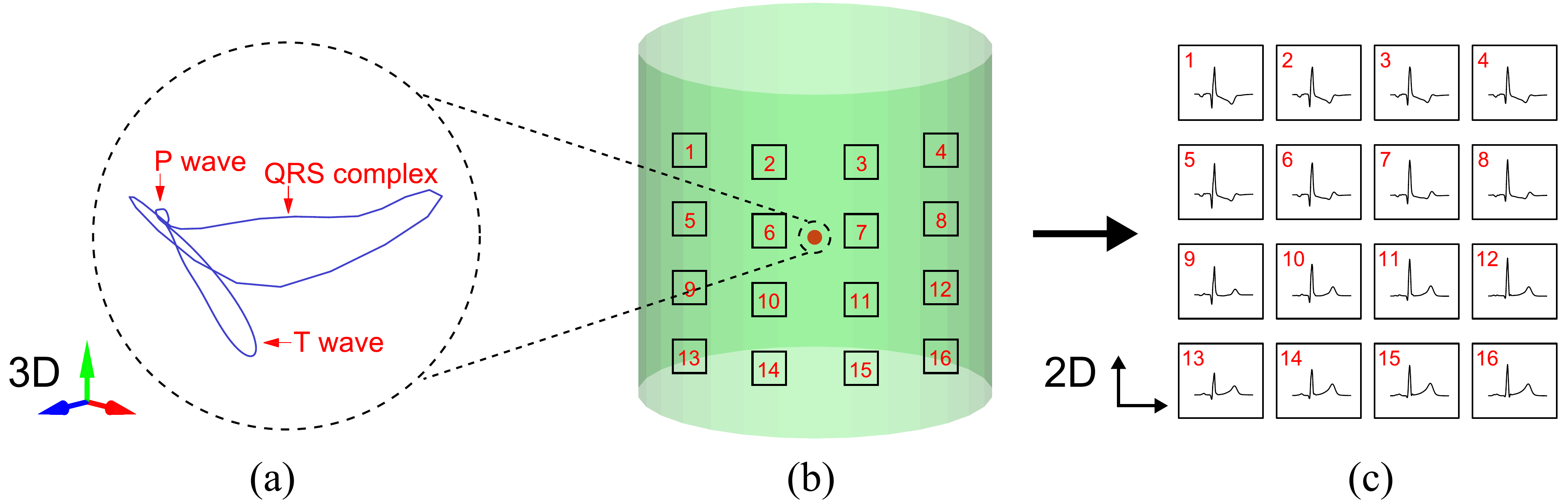}
	\caption{Steps in generating a dynamic abdominal ECG mixture as implemented in the \textit{fecgsyn} toolbox \highlightcolor{(a)} Vector loop representing the time-varying source model \highlightcolor{(b)} Homogeneous volume conductor model representing the maternal-fetal anatomy \highlightcolor{(c)} Resulting surface potentials where numbered squares indicate sensor positions. Note the P wave, QRS complex and T wave are oriented differently in 3D space.}
	\label{modelling_overview}
\end{figure*}

\section{Background}

To date, the accuracy of NI-FECG extraction algorithms on clinical data has primarily been assessed using two types of reference signal: 1) Data recorded from an intrauterine fetal scalp electrode (FSE) \cite{Graatsma_Fetalelectrocardiographyfeasibility_2009,Clifford_Clinicallyaccuratefetal_2011,Andreotti_RobustfetalECG_2014} or 2) Expert annotations of the NI-FECG \cite{Hoyer_Monitoringfetalmaturation_2017}. While these approaches are appropriate for validating the temporal accuracy of individual event detections (e.g. QRS complex locations), they are unsuitable for verifying the 3D characteristics of fetal cardiac signals as the FSE records a 1D projection at the placement site and expert annotations rely on the assumption that signal morphology has not been altered by volume conductor effects prior to inspection. 

In an attempt to provide a consistent reference signal against which NI-FECG extraction algorithms can be benchmarked, ongoing work in simulating dynamic maternal-fetal ECG mixtures from a synthetic cardiac source has been completed by Sameni et. al\cite{Sameni_MultichannelECGNoise_2007}, Behar et al. \cite{Behar_ECGsimulatorgenerating_2014} and Andreotti et al. \cite{Andreotti_opensourceframeworkstresstesting_2016} to produce the open-source \textit{fecgsyn} toolbox. This MATLAB based toolbox enables the simulation of maternal-fetal ECG mixtures that incorporate beat-to-beat variability, fetal movement and realistic noise, allowing benchmarks in a range of pathological scenarios. These simulations are important to investigate NI-FECG extraction accuracy in cases where the NI-FECG cannot be visually annotated or placement of the FSE is infeasible, such as in early gestation. However, as the \textit{fecgsyn} toolbox utilizes a homogeneous volume conductor model, it is important to acknowledge that such simulations may not capture all properties of the true system and could provide false confidence in extraction accuracy when applied to real data.

To simulate the propagation of cardiac electrical activity to potentials on the maternal abdomen, there are two key components that must be chosen: 1) The source model or electrical approximation of cardiac activity and 2) The volume conductor model through which this electrical activity propagates. An overview of the simulation process currently used in the \textit{fecgsyn} toolbox is shown in \figref{modelling_overview} which describes: a) vector loop representing the time-varying source model, b) homogeneous volume conductor model representing the maternal-fetal anatomy and c) resulting surface potentials where numbered squares indicate sensor positions on the volume conductor.

\subsection{Source model}
There are a number of source models available to approximate cardiac electrical activity, ranging from the point dipole to the equivalent double layer (EDL). For an in-depth discussion of these models and their physiological basis, the reader is referred to \cite{Malmivuo_Bioelectromagnetismprinciplesapplications_1995}. The source model used in the \textit{fecgsyn} toolbox is the point dipole due to its straightforward volume conductor solution and ability to approximate 80\% to 90\% of the power in observed surface potentials \cite{Malmivuo_Bioelectromagnetismprinciplesapplications_1995,van2002beyond}. The time-varying potential vector $\phi(t)$ produced by a point dipole for a chosen set of sensor positions in the \textit{fecgsyn} toolbox is given by:
\vspace*{-0.2cm}

\begin{equation}
\phi(t) = H(t)\cdot R(t)\cdot d(t)
\label{potential_single}
\end{equation}

\noindent where for $n$ sensor positions, $d(t)$ is a $3\times1$ vector representing the point dipole, $R(t)$ is a $3\times3$ rotation matrix of the point dipole and $H(t)$ is a $n\times3$ lead field matrix. The lead field matrix $H(t)$ represents the transformation from source vector to scalar potential at each sensor position as given by the volume conductor model (\textit{see Section \rom{2}B}). Both $H(t)$ and $R(t)$ may be time-varying when the cardiac source is in motion. Combining the maternal and fetal cardiac sources with noise results in:
\vspace*{-0.3cm}

\begin{equation}
\phi(t) = H_m(t)\cdot R_m(t)\cdot d_m(t)+H_f(t)\cdot R_f(t)\cdot d_f(t)+w(t)
\label{potential_multiple}
\vspace*{0.1cm}
\end{equation}

\noindent where the subscripts $m$ and $f$ indicate the respective maternal and fetal parameters and $w(t)$ is noise.

\subsection{Volume conductor model}
To calculate the potential distribution generated by electrical sources within the human body, a model of its electrical properties must be defined. The volume conductor model used in the \textit{fecgsyn} toolbox to calculate $H(t)$  is summarized as follows with further detail available in \cite{Sameni_MultichannelECGNoise_2007} and \cite{Geselowitz_theoryelectrocardiogram_1989}. Considering the human body as a linear, resistive volume, the quasi-static potential $\phi$ at a sensor position external to any current source can be described by the Poisson equation:

\begin{equation}
\nabla\cdot\sigma\nabla\phi = \nabla\cdot \textbf{J}^i
\label{poisson}
\end{equation}

\noindent where $\textbf{J}^i$ is the impressed current density and $\sigma$ is conductivity. Assuming an infinite, homogeneous volume, a solution to \eqref{poisson} is given by:
\vspace*{-0.15cm}

\begin{equation}
\phi(r) - \phi_0 = \frac{1}{4\pi\sigma}\iiint\limits_V \frac{\nabla\cdot\textbf{J}^i}{\lvert r\rvert}\,dV
\label{poisson_solution}
\end{equation}

\noindent where $r$ is the vector from source to sensor position and $\phi_0$ a chosen reference potential. For a single time-varying source, \eqref{poisson_solution} has a solution:
\vspace*{-0.12cm}

\begin{equation}
\phi(t) - \phi_0 = \frac{r(t)}{4\pi\sigma\lvert r(t)\rvert^3}\cdot d(t) 
\label{poisson_solution_analytic}
\end{equation}

\noindent where $d(t)$ is the time-varying vector representing the point dipole and $r(t)$ is the time-varying $r$ as previously defined. Equation \eqref{poisson_solution_analytic} expresses the solution for the elements of $H(t)$ as used in the \textit{fecgsyn} toolbox with the rotation matrix $R(t)$ used to modify the overall dipole orientation. However, this solution is only valid for a volume conductor comprised of an infinite homogeneous medium. As an inhomogeneous volume conductor affects dipole sources depending on their specific position and orientation, the proposal by Behar et al. \cite{Behar_ECGsimulatorgenerating_2014} to model vernix caseosa by reducing the signal-to-noise ratio (SNR) of the fetal cardiac source will not accurately reflect morphological changes introduced by an asymmetric volume conductor.
\vspace*{-0.2cm}
\section{Method}
\subsection{Asymmetric volume conductor modeling}
In this work, we extend the \textit{fecgsyn} toolbox by replacing $H(t)$ with a lead field matrix computed using the finite element method (FEM). The FEM operates by discretizing the volume conductor into smaller elements with defined electromagnetic properties, enabling the computation of electric potential and magnetic field distribution in complex geometric structures. This technique has been used previously in the field of electroencephalography (EEG) where inhomogeneities such as the highly non-conductive skull have a significant effect on EEG measurements \cite{Chauveau_Effectsskullthickness_2004}. Several open-source EEG toolboxes with volume conductor modeling capability have been released to study these effects including the Neuroelectromagnetic Forward Head Modeling Toolbox (NFT) \cite{Acar_NeuroelectromagneticForwardHead_2010}, Brainstorm \cite{Tadel_BrainstormUserFriendlyApplication_2011} and FieldTrip \cite{Oostenveld_FieldTripOpenSource_2011}. In this work, we utilize the FieldTrip-Simbio FEM pipeline developed by Vorwerk et al. \cite{Vorwerk_FieldTripSimBiopipelineEEG_2018}. This pipeline has been chosen due to its numerical accuracy \cite{Vorwerk_ComparisonBoundaryElement_2012} and straightforward integration into the MATLAB environment. For further reference on its implementation, the reader is referred to \cite{Vorwerk_Newfiniteelement_2016}.

To describe the maternal-fetal anatomy, we use a model containing 4 tissue types consisting of fetus, vernix caseosa, amniotic fluid and maternal abdomen as proposed by Stinstra et al. \cite{Stinstra_InfluenceFetoabdominalTissues_2002}. We assume the maternal-fetal anatomy $\Omega$ can be decomposed into a finite number of compartments $\Omega_1,\Omega_2,\ldots,\Omega_n$ each containing a tissue type of a single resistive conductivity $\sigma_i$. Adjacent compartments must share a common boundary and there may be disjoint compartments containing the same tissue type within the overall domain. The electrical conductivities used for each tissue type in our model are taken from \cite{Stinstra_InfluenceFetoabdominalTissues_2002} and are given in \tableref{conductivity}.

\begin{table}[H]
	\renewcommand{\arraystretch}{1.2}
	\caption{Tissue Conductivities. Data from \cite{Stinstra_InfluenceFetoabdominalTissues_2002}}
	\label{conductivity}
	\centering
	\begin{tabular}{C{0.22\textwidth} C{0.22\textwidth}} 
		\hline
		\hline
		\bfseries Tissue & \bfseries Conductivity (S/m) \\
		\hline
		Fetus & 0.5\\
		Vernix caseosa & 10\textsuperscript{-5}\\
		Amniotic fluid & 1.6 \\
		Maternal body & 0.2\\
		\hline
		\hline
	\end{tabular}
\end{table}

\begin{figure}[h]
	\centering
	\includegraphics[width=0.485\textwidth]{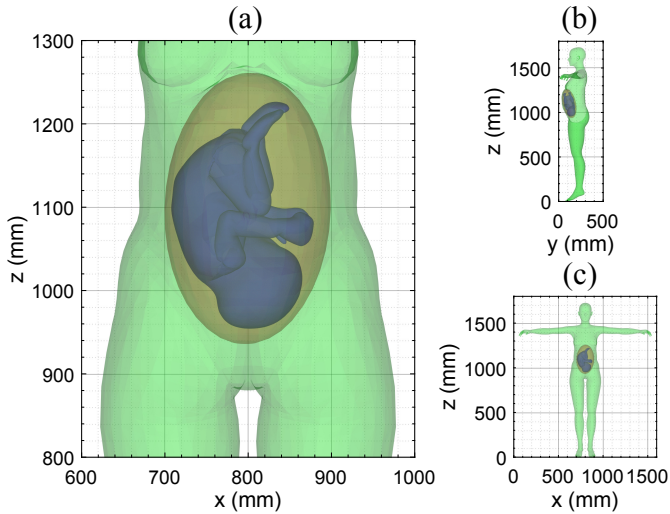} 
	\caption{32 weeks GA maternal-fetal model with fetus in right occiput transverse (ROT) presentation \highlightcolor{(a)} Abdominal view \highlightcolor{(b)} Sagittal view \highlightcolor{(c)} Coronal view where \protect\mycbox{shadedFetus} = fetus, \protect\mycbox{shadedAmniotic} = amniotic fluid and \protect\mycbox{shadedAbdomen} = maternal body}
	\label{maternal_fetal_model}
\end{figure}

To generate a finite element model based on realistic maternal-fetal geometry, we utilize the publically available Fetus and Mother Numerical Models (FEMONUM) as described in Bibin et al. \cite{Bibin_Wholebodypregnantwoman_2010} and Dahdouh et al. \cite{Dahdouh_comprehensivetoolimagebased_2014}. These models were created using a combination of 3D ultrasound and magnetic resonance imaging (MRI) data from women with pregnancies ranging from 8 to 34 weeks GA. Specifically we use the \textit{Victoria-TiGroFetus-32weeks} model composed of the maternal body model Victoria (provided by DAZ 3D Studio,\;\url{www.daz3d.com}) and 32 weeks GA model which discretizes the fetus, amniotic fluid and maternal body into 3 triangulated compartments of 24901, 382 and 11149 nodes respectively as shown in \figref{maternal_fetal_model}.

Before use, all models were processed using the open-source MeshLab toolbox developed by Cignoni et al. \cite{Cignoni_MeshLabOpenSourceMesh_2008} to achieve the following: 1) Remove self-intersections, 2) Smooth and resample each compartment to create a single closed surface and 3) Compute the constrained Delaunay triangulation (CDT) of the resulting surface. 

\begin{figure}
	\centering
	\includegraphics[width=0.48\textwidth]{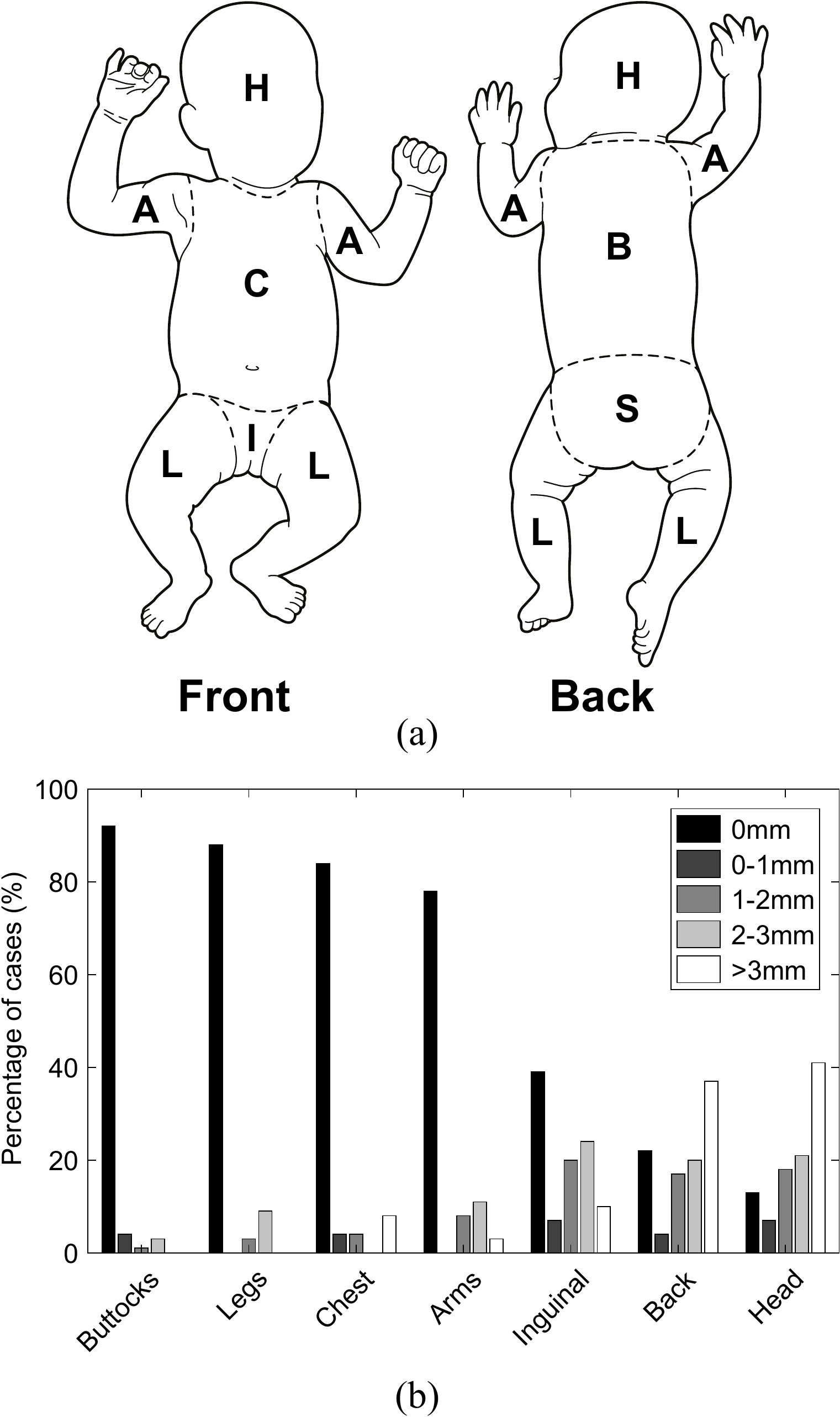}
	\caption{\highlightcolor{(a)} Fetal body maps\protect\footnotemark\:where A = Arms, B = Back, C = Chest, H = Head, I = Inguinal, L = Legs, S = Buttocks. \highlightcolor{(b)} Distribution of vernix caseosa in each region as identified from  fetal body maps. Total cases (n=100) where cases $<$37 weeks GA (n=30), 37.1$-$40.9 weeks GA (n=59) and $>=$ 41 weeks GA (n=11). Data from \cite{Archana_ClinicalStudySurface_2008}}
	\label{fetal_body_map}
\end{figure}

To describe the asymmetric distribution of vernix caseosa we propose a model that delineates the fetal body into 7 regions as defined by the fetal body maps in \figrefmult{fetal_body_map}{a}. The physiological basis for this model comes from: 1) observations of vernix caseosa asymmetry in these regions as presented by Akiba \cite{Akiba_Studiesbiologicalactions_1955} and Visscher et al. \cite{Visscher_VernixCaseosaNeonatal_2005} and 2) results from Archana \cite{Archana_ClinicalStudySurface_2008} which classify vernix caseosa thickness in these regions into five ranges ($0\text{mm},\;0-1\text{mm},\;1-2\text{mm},\;2-3\text{mm},\;>3\text{mm}$) as shown in \figrefmult{fetal_body_map}{b}. The correlation between vernix caseosa thickness in each region and breakdown by gestational age was not reported by Archana, therefore these results should not be interpreted as the typical vernix caseosa distribution for an individual but instead serve as an indicative range of values to be studied.\footnotetext{LifeART image copyright 2000 Wolters Kluwer Health, Inc.- Lippincott Williams \& Wilkin. All rights reserved.}

To investigate the range of observed vernix caseosa thicknesses, we manually divide the processed 32 weeks GA fetal model into the 7 regions as defined in \figrefmult{fetal_body_map}{a}. Compartments of vernix caseosa are generated by selecting a group of connected triangles $S_1$ representing a desired region and translating its inner nodes along the vertex normals to create a new surface $S_2$. Following this, a triangulated surface $S_3$ is created joining the exterior nodes of $S_1$ and $S_2$ to define the vernix caseosa compartment bounded by $S_1 \cup S_2 \cup S_3$. This process is repeated until all desired vernix caseosa compartments have been created. It is important to note the true distribution of vernix caseosa may be more granular than these 7 regions, however this model will help identify regions which have a significant effect on NI-FECG extraction accuracy.

\begin{figure}[h]
	\centering
	\includegraphics[width=0.49\textwidth]{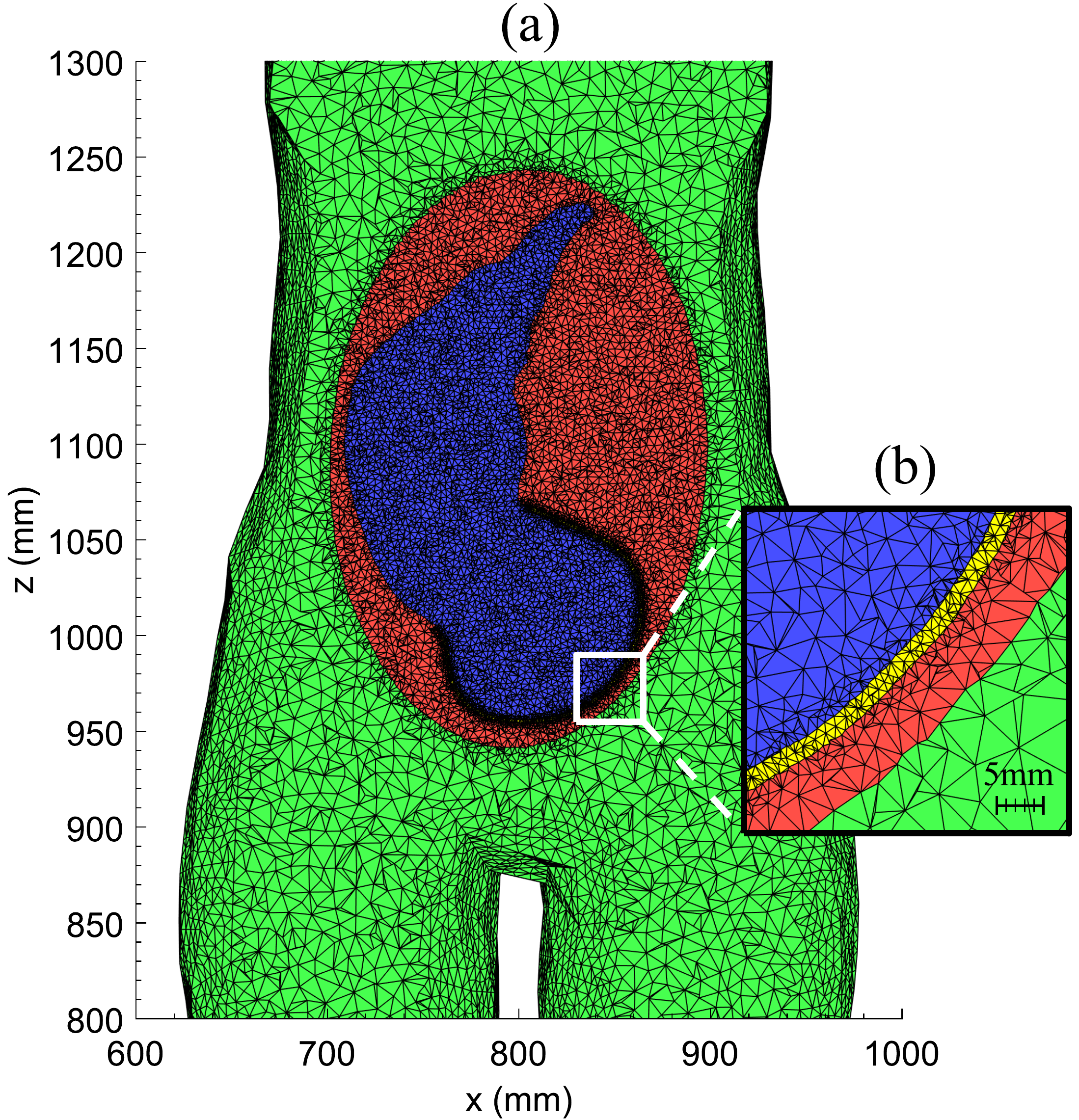}
	\caption{\highlightcolor{(a)} Cross section  (y $>$ 127mm) of tetrahedral finite element model with vernix caseosa generated at 2mm thickness in the head region \mbox{\highlightcolor{(b)} Inset view} showing fine discretization in the vernix caseosa compartment. Tissue types are color coded as follows: \protect\mycbox{fetus} = fetus, \protect\mycbox{vernix} = vernix caseosa, \protect\mycbox{amniotic} = amniotic fluid and \protect\mycbox{abdomen} = maternal body}
	\label{fem_model}
\end{figure}

To facilitate finite element model generation from the defined surfaces, the compartments of fetus, vernix caseosa, amniotic fluid and maternal body are combined to form a triangulated piecewise linear complex (PLC) \cite{Miller_Controlvolumemeshes_1998}. From the triangular PLC, a tetrahedral finite element model can be generated via Constrained Delaunay tetrahedralization. In this work, this process is achieved using the open-source TetGen tool (v1.5.1) developed by Si \cite{Si_TetGenDelaunayBasedQuality_2015} invoked through the iso2mesh MATLAB toolbox (v1.8.0) developed by Fang et al.  \cite{Fang_Tetrahedralmeshgeneration_2009}. TetGen allows the user to set a maximum element volume for the discretization of each compartment, important for the thin layer of vernix caseosa to ensure an accurate FEM solution. An example tetrahedral finite element model generated using this process is shown in \figref{fem_model}

\vspace{-0.5cm}
\subsection{Source parameters}

Each cardiac source within this model must be assigned three parameters: 1) position of the source, 2) orientation of the source representing the \textit{electrical axis} of the heart and 3) source vector loop, otherwise known as the \textit{vectorcardiogram}, representing the 3D path traced by the source model. For the maternal heart these parameters are affected by inter-individual variations and physiological changes in the heart during pregnancy as well as fetal presentation and fetal size due to the displacement of internal organs \cite{Fowler_AnatomicElectrocardiographicPosition_1951,Bacharova_LeftVentricularHypertrophy_2014,Gultekin_whatextentpresentation_2016}. As the scope of investigating these parameters is beyond this work, the default position and orientation of the maternal source in our model are determined based on the mean values for a normal adult heart as reported by \mbox{Nousiainen et al. \cite{Nousiainen_Comparativestudynormal_1986}} with the ability to select from nine different vectorcardiograms as provided in the \textit{fecgsyn} toolbox.

For the fetal source, in addition to inter-individual variations, the ratio of right ventricular to left ventricular weight varies throughout gestation resulting in a shift in electrical axis due to the increased thickness of muscle fibre \cite{Emery_WeightVentriclesLater_1960}. While a recent study \cite{Verdurmen_Orientationelectricalheart_2016} has recorded preliminary results for the normal fetal electrical axis, they are not suitable for use in this work due to their small sample size. Instead, the default position of the fetal source in our model is based on the normal position for a healthy fetus (13-40 weeks GA) as reported in \cite{Comstock_NormalFetalHeart_1987} with the same anatomic orientation as the maternal source adjusted for right ventricular deviation as observed in early neonates \cite{DePasquale_electrocardiogramventriculargradient_1963}. A summary of the default source position, orientation, vectorcardiogram and relevant \textit{fecgsyn} parameters utilized in our model are shown in \tableref{source_parameters}.

\begin{table}[H]
	\renewcommand{\arraystretch}{1.2}
	\caption{Source Model Parameters}
	\label{source_parameters}
	\centering
	\begin{tabular}{m{0.07\textwidth} m{0.24\textwidth} m{0.1\textwidth}} 
		\hline
		\hline
		\bfseries Parameter & \bfseries Description & \bfseries Value \\
		\hline
		fectb & Fetal ectopic beats & 0\\
		fheart & Fetal source position ($x,y,z$) & ($772,125,1087$)\\
		fhr & Fetal heart rate & 150\\
		ftraj & Fetal movement trajectory & none\\
		ftypeacc & Fetal heart rate acceleration & none\\
		fvcg & Fetal vectorcardiogram & 1\\
		mectb & Maternal ectopic beats & 0\\
		mheart & Maternal source position ($x,y,z$) & ($814,165,1315$)\\
		mhr & Maternal heart rate & 80\\
		mtraj & Maternal movement trajectory & none\\
		mtypeacc & Maternal heart rate acceleration & none\\
		mvcg & Maternal vectorcardiogram & 1\\
		posdev & Heart position deviation & 0 \\
		$R_f$ & Fetal source rotation ($\theta_x,\theta_y,\theta_z$)& (2.8,1.6,3)\\
		$R_m$ & Maternal source rotation ($\theta_x,\theta_y,\theta_z$)& (1.5,0,0)\\
		\hline
		\hline
	\end{tabular}
\end{table}

\subsection{Metrics}
Assessment of volume conductor effects is performed by comparing lead field matrix components ($x,y,z$) and surface potentials at selected nodes in each model using the relative difference measure ($RDM^*$) as proposed by Meijs et al. \cite{Meijs_numericalaccuracyboundary_1989} and the logarithmic magnitude error ($lnMAG$) as proposed by G\"{u}llmar et al. \cite{Gullmar_Influenceanisotropicelectrical_2010}:
\vspace*{-0.1cm}

\begin{equation*}
RDM^*=\norm[\Big]{\frac{x_a}{\norm{x_a}_2}-\frac{x_b}{\norm{x_b}_2}}_2\quad
lnMAG=\text{ln}\Big(\frac{\norm{x_a}_2}{\norm{x_b}_2}\Big)
\end{equation*}

\begin{figure}
	\centering
	\hspace*{-0.4cm}   
	\includegraphics[width=0.47\textwidth]{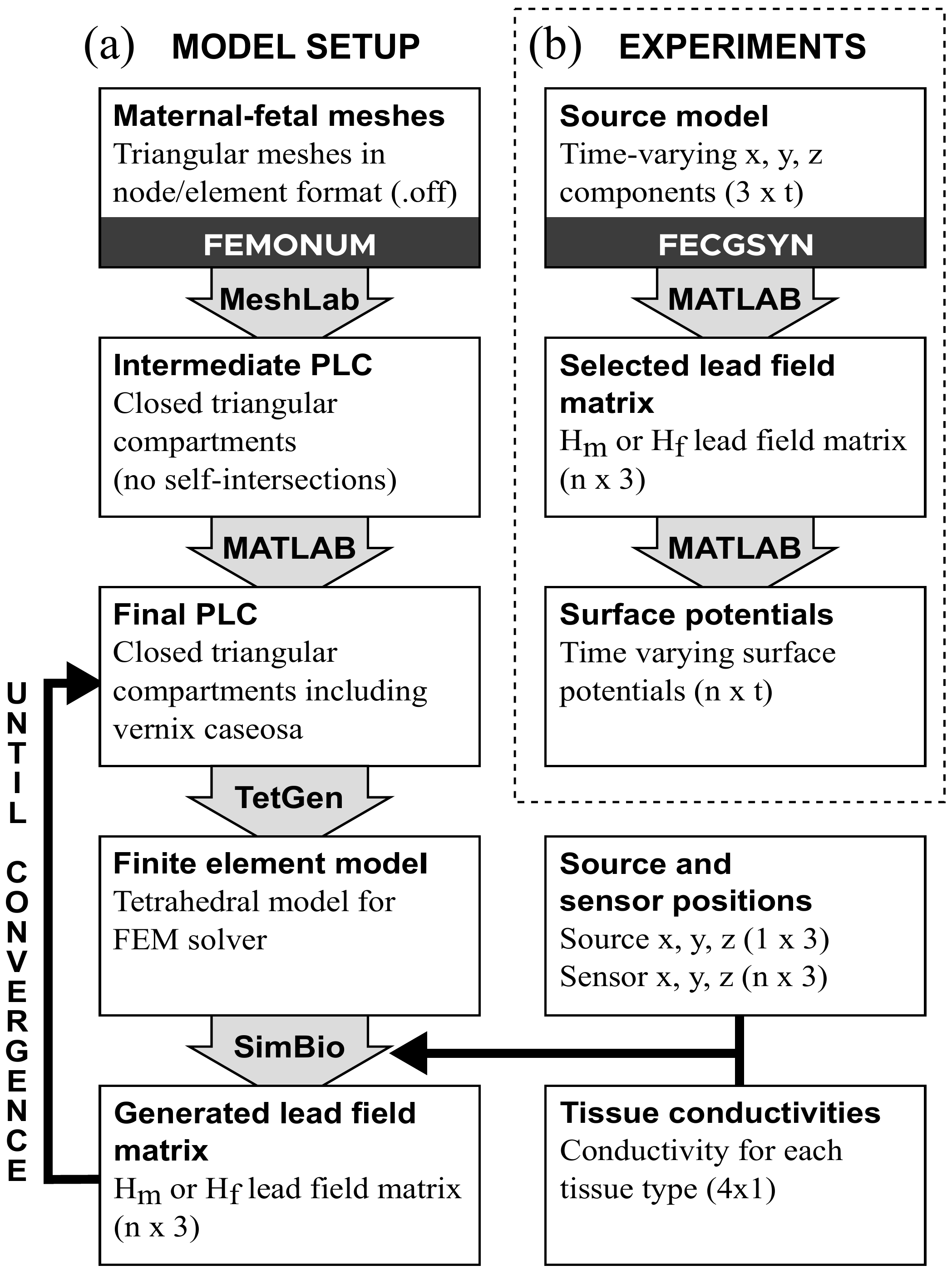}	
	\caption{Developed process for generating surface potentials in an asymmetric volume conductor model. Dark gray boxes indicate data from the labeled source. Light gray arrows indicate a processing step using the labeled tool. Black arrows indicate data input to the target. \highlightcolor{(a)} indicates steps utilized for model setup and \highlightcolor{(b)} indicates steps utilized for the experiments.}
	\label{overall_process}
\end{figure}
\vspace*{0.1cm}
\noindent where for $n$ nodes, $x_a$ and $x_b$ are $n\times1$ vectors for inputs $a$ and $b$ respectively. $RDM^*$ indicates the difference in distribution patterns, bounded by 0 for identical inputs and 2 for $x_a = -x_b$. $lnMAG$ indicates the difference in overall magnitude where a value of 0 indicates identical inputs and positive or negative deviation indicates relative change in magnitude. $lnMAG$ is preferable to a direct magnitude ratio as it is symmetric about $ln(1)$, enabling one-to-one comparison between positive and negative differences. As we aim to quantify effects for sources of varying orientation, we calculate $RDM^*$ and $lnMAG$ for the lead field matrix as 3x1 vectors $[RDM^*_x,\; RDM^*_y,\;RDM^*_z]$ and $[lnMAG_x,\; lnMAG_y,\;lnMAG_z]$ indicating the value along each coordinate axis. Additionally, as the T/QRS ratio has been proposed for identifying signs of fetal distress, we define the T/QRS ratio error in each volume conductor model as:
\vspace*{0.3cm}
\begin{equation*}
e_{TQRS}=\frac{TQRS_{a} - TQRS_{h}}{TQRS_{h}}
\end{equation*}\\
\noindent where $TQRS_{a}$ is the T/QRS ratio in the asymmetric volume conductor model and $TQRS_{h}$ is the T/QRS ratio in the homogeneous volume conductor model.

\subsection{Model Setup}

\begin{figure}
	\centering
	\includegraphics[width=0.47\textwidth]{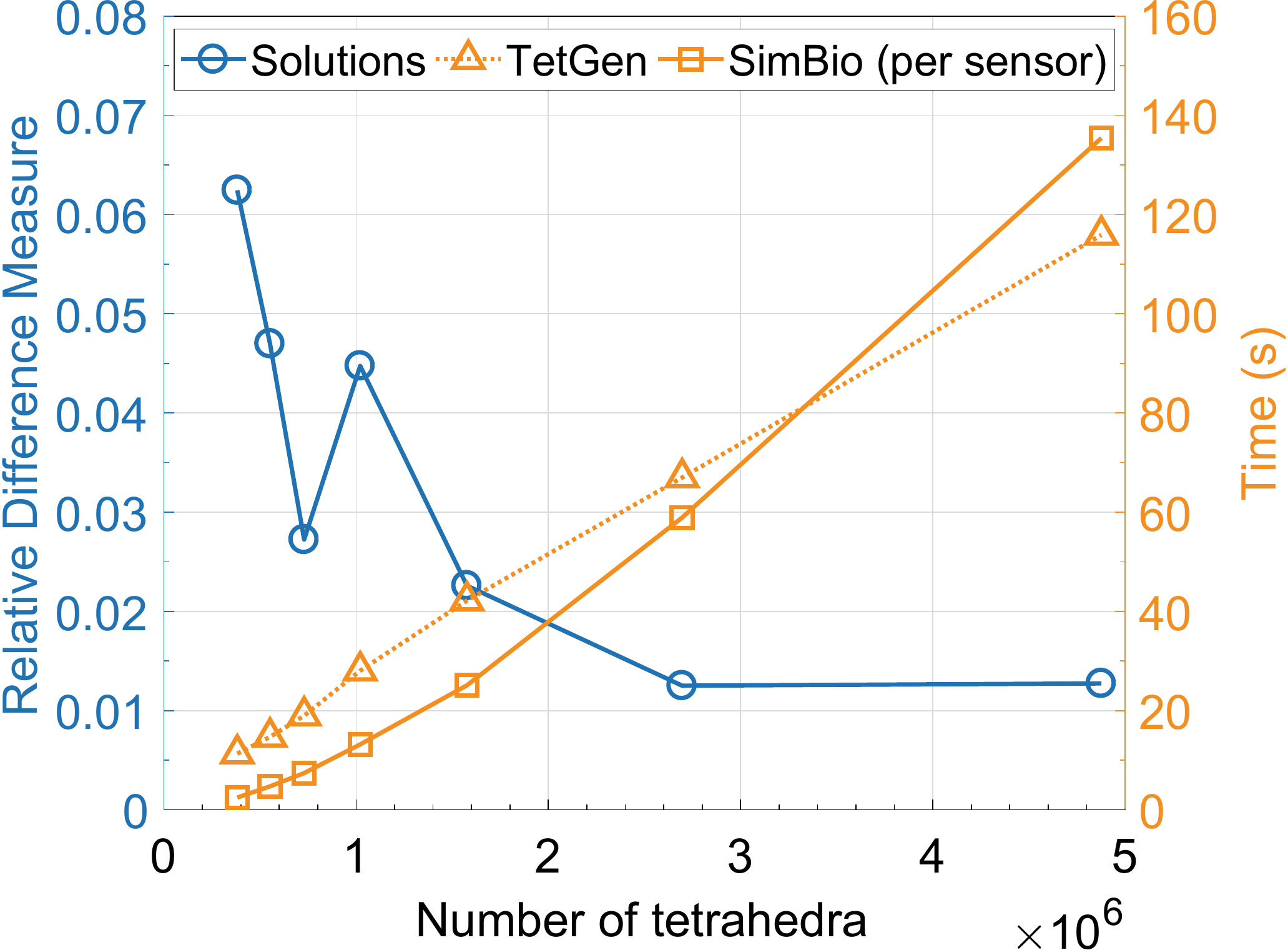}
	\caption{Model refinement process showing lead field matrix convergence for the Back 1mm model as maximum element volume per compartment is halved, approximately doubling the number of tetrahedra at each step. \protect\includegraphics[width=0.02\textwidth]{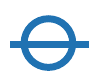} shows a converging relative difference measure ($\norm{RDM^*}_2$) calculated at each step compared to the subsequent solution, \protect\includegraphics[width=0.02\textwidth]{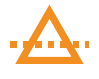} shows time to compute tetrahedralization via TetGen and \protect\includegraphics[width=0.02\textwidth]{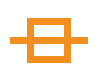} shows time to compute lead field matrix via SimBio per sensor. Subsequent reference solution for the final data point is not shown.}
	\label{model_refinement}
\end{figure}

To quantify the effects of volume conductor asymmetry on NI-FECG extraction, we compute the fetal lead matrix ($H_f$) at all maternal body surface nodes between $750 \leq z \leq 1350$ ($n=682$) for visualisation purposes and define a subset of nodes between $950 < z < 1200$ ($n=266$) as the \textit{abdominal sensors} for numerical evaluation. To ensure minimal numerical error in the chosen finite element discretization, we perform a refinement process using a model with vernix caseosa generated at 1mm in the Back region to determine the required TetGen parameters. This process involves repeatedly completing the TetGen and SimBio steps as shown in \figref{overall_process} and iteratively halving the maximum element volume per compartment until a defined level of convergence is obtained. For this work, refinement was performed until $\norm{RDM^*}_2$ of the \textit{abdominal sensors} lead field matrix compared with the subsequent solution was below 0.02 for two consecutive iterations as shown in \figref{model_refinement}. All finite element models in this work were generated using the parameters identified via this process and compared to a reference solution with halved maximum element volume per compartment showing minimal numerical error in $RDM^*$ and $lnMAG$ (results in Appendix \tableref{model_refinement_results}).

\subsection{Experiments}
Using the developed process, this work aims to characterise morphological changes caused by volume conductor asymmetry including spatially variable distributions of vernix caseosa in the 7 fetal body regions as defined in \figrefmult{fetal_body_map}{a}. Based on the data shown in \figrefmult{fetal_body_map}{b}, it would be ideal to investigate models where vernix caseosa thickness in each region takes one of four values: 0mm, 1mm, 2mm or 3mm. However, as this space represents $16384\;(4^{7})$ possible combinations, we must define a subset of this space to make analysis feasible.

\subsubsection{Region Assessment}

In the first experiment, we quantify regions which have the most significant impact by analysing a subset of volume conductor models consisting of the following 16 scenarios: \mbox{1) homogeneous} volume conductor model as defined by Equation (5) with conductivity set to that of the maternal abdomen (0.2 S/m), 2) finite element model with no vernix caseosa and 3-16) finite element models consisting of vernix caseosa isolated in each region at 1mm and 3mm thickness (minimum and maximum non-zero values). Using the generated lead field matrices, we calculate pairwise $RDM^*$ and $lnMAG$ along each coordinate axis for each of the 16 volume conductor models. From these results, the three regions with the greatest $\norm{RDM^*}_2$ and $\norm{lnMAG}_2$ compared to the No Vernix model are selected for further analyis. The No Vernix model is used for comparison as it characterizes regional effects more precisely compared to shared differences with respect to the homogeneous model. 

\subsubsection{Signal Morphology}
In the second experiment, we generate asymmetric volume conductor models in the three identified regions and assess changes to NI-FECG morphology using two methods: 1) visualizing surface potentials produced by dipoles along each coordinate axis and 2) calculating $e_{TQRS}$ and $RDM^*$ at 6 sensor positions on the maternal abdomen compared to a homogeneous volume conductor model. This is achieved by simulating a 300ms vectorcardiogram containing one fetal cardiac cycle using the parameters defined in \tableref{source_parameters} and calculating the NI-FECG waveform at each sensor per volume conductor model. As current benchmarks using \textit{fecgsyn} assume that accurately extracting this component from an abdominal mixture in a homogeneous volume conductor model indicates correct identification of the underlying source, quantifying the observed error due to volume conductor effects demonstrates the potential for estimation error in real data using this approach.

To calculate the T/QRS ratio for each NI-FECG waveform, the QRS complex amplitude is measured as the difference between the minimum and maximum values in the QRS complex detection zone (defined as 80 to 120ms) while the T wave amplitude is measured as the maximum absolute value in the T wave detection zone (defined as 180 to 220ms) with respect to the isoelectric line (defined as the value at 300ms). This process is necessary as the exact sample which represents the minimum/maximum QRS complex amplitude and maximum T wave amplitude may shift slightly depending on the chosen sensor position and volume conductor model.

\section{Results}
This section presents numerical results for the experiments described. All simulations were performed in MATLAB R2017a on the University of Melbourne's High Performance Computing system \cite{MEADE_SpartanHPCCloudHybrid_2017} comprised of 100 Linux computing nodes each with an eight-core 2.6Ghz Intel Xeon CPU and 62GB RAM with each simulation running on a single node. 

\begin{figure*}
	\centering
	\vspace*{-0.2cm}
	\includegraphics[width=1\textwidth]{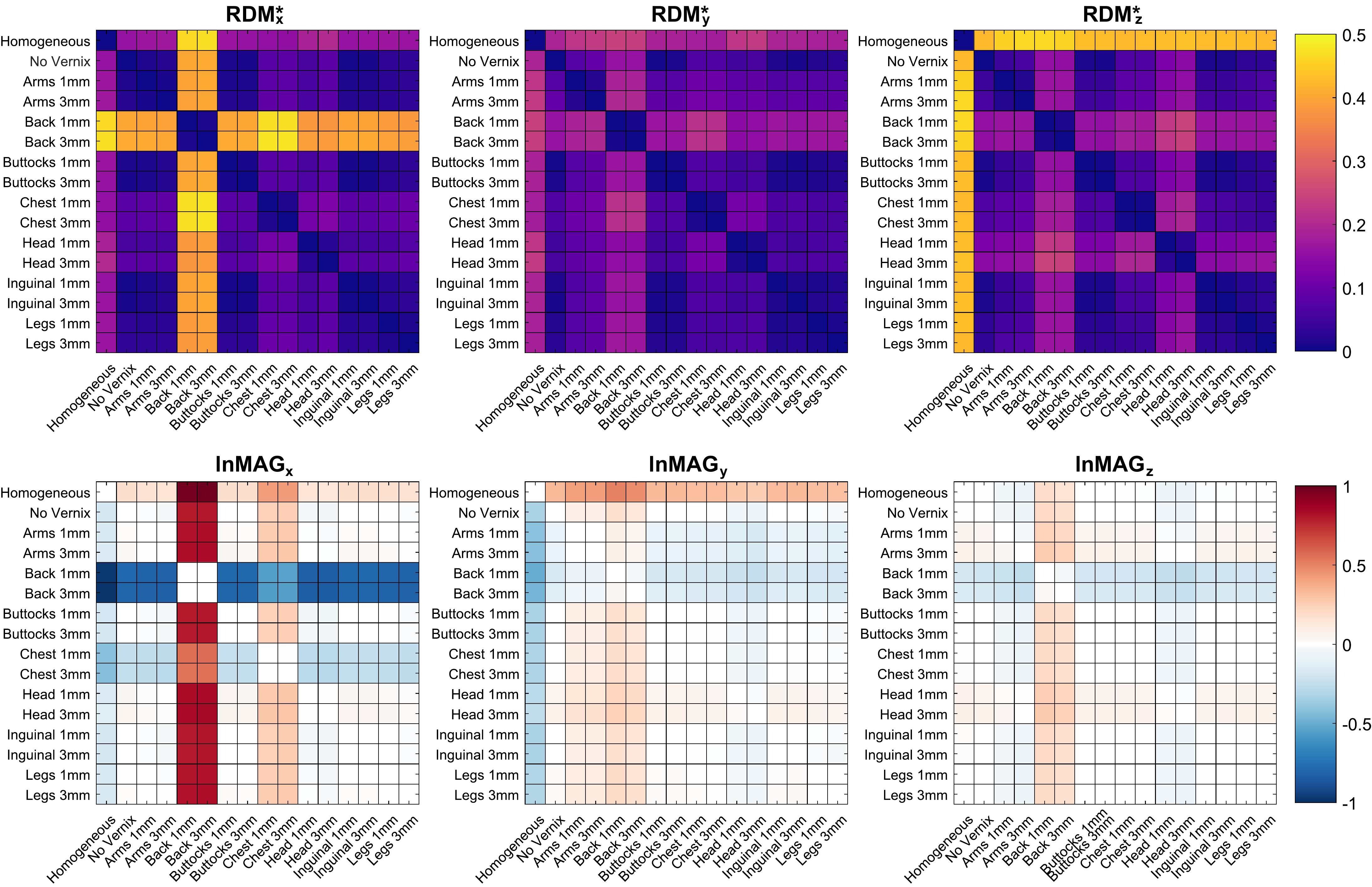}
	\caption{Region assessment results showing pairwise comparison of $RDM^*$ and $lnMAG$ for $x,y,z$ components of the \textit{abdominal sensors} lead field matrix. Each colored square indicates the metric defined in Section IIC between pairs of models where rows = $x_a$ and columns = $x_b$. Increasing $RDM*$ represents greater difference in distribution patterns and positive or negative $lnMAG$ represents relative change in magnitude.}
	\label{heatmap}
\end{figure*}

\subsection{Region Assessment}

Using the described volume conductor models, pairwise $RDM^*$ and $lnMAG$ were computed for the \textit{abdominal sensors} lead field matrix as shown in \figref{heatmap}. Each heat map indicates the relative change in surface potentials between models in terms of distribution patterns ($RDM^*$) and overall magnitude ($lnMAG$) as generated by components of the fetal source along each coordinate axis.

From these results, we observe that the Back models demonstrate the largest change in $RDM^*_x$ compared to the homogeneous model followed by the Head, Arms and Chest models. Differences in $RDM^*_y$ are less pronounced with the Back, Arms and Head models showing marginally greater change, while for $RDM^*_z$ all models show significant change compared to the homogeneous model. For all models with vernix caseosa, it can be observed that varying vernix caseosa thickness from 1mm to 3mm in each region has minimal effect on $RDM^*$ along all coordinate axes.

For $lnMAG_x$, all models show a decrease compared to the homogeneous model with the Back and Chest models demonstrating the greatest change. Interestingly, while $lnMAG_x$ for the Head models is decreased compared to the homogeneous model, it is marginally increased compared to all other models. For $lnMAG_y$, all models exhibit a large decrease compared to the homogeneous model with the Head model again showing a small increase in $lnMAG_y$ compared to non-homogeneous models. Most models show minimal change in $lnMAG_z$ compared to the homogeneous model except for the Arms and Head models which marginally increase and the Back models which marginally decrease. Similar to $RDM^*$, varying the thickness of vernix caseosa from 1mm to 3mm in each region has minimal effect on $lnMAG$ along all coordinate axes.

\begin{figure}[H]
	\centering
	\includegraphics[width=0.48\textwidth]{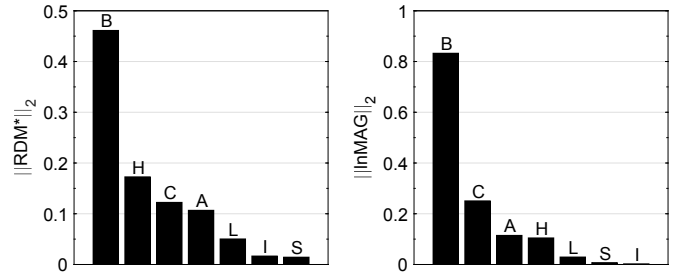}	
	\caption{$\norm{RDM^*}_2$ and $\norm{lnMAG}_2$ rankings for 3mm models compared to No Vernix model where A = Arms, B = Back, C = Chest, H = Head, I = Inguinal, L = Legs, S = Buttocks}
	\label{ranking_results}
\end{figure}

To quantify the three regions of greatest impact, we calculate $\norm{RDM^*}_2$ and $\norm{lnMAG}_2$ rankings for the 3mm models compared to the No Vernix model as shown in \figref{ranking_results}. Ranking is performed using the 3mm models as overall they demonstrate marginally greater impact compared to the 1mm models. From these rankings, the Back 3mm model shows the greatest change in $\norm{RDM^*}_2$ followed by the Head 3mm and Chest 3mm models. For $\norm{lnMAG}_2$, the Back 3mm model shows the greatest change, followed by the Chest 3mm and Arms 3mm models. As this work is primarily concerned with morphological changes indicated by greater $RDM^*$, the three models selected for further analysis are the Back 3mm, Chest 3mm and Head 3mm models.
\vspace{-0.1cm}
\subsection{Signal Morphology}
\begin{figure*}
	\centering
	\includegraphics[width=1\textwidth]{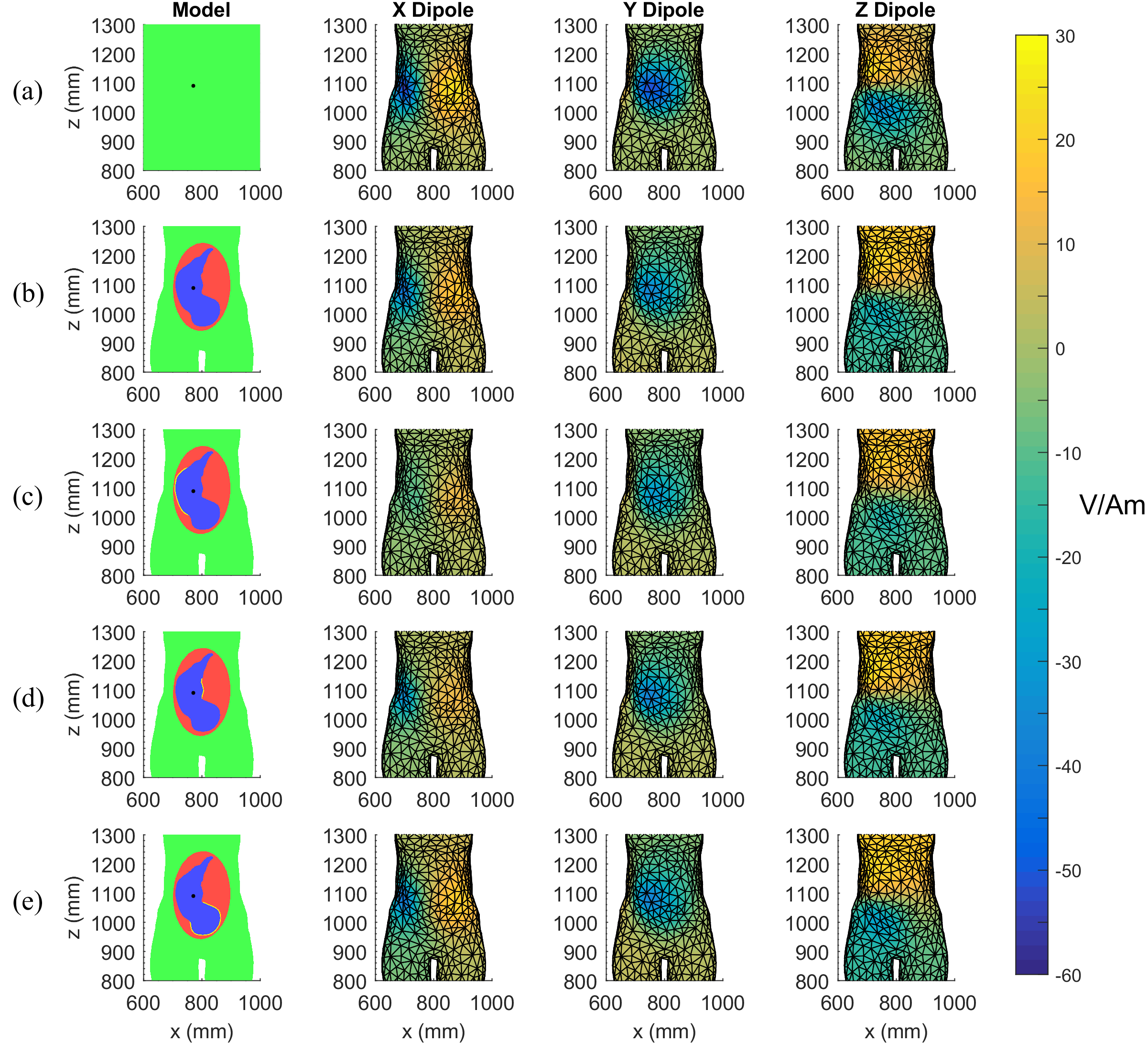}
	\caption{Cross section (y $>$ 127mm) of selected volume conductor models and resulting surface potentials produced by dipoles along each co-ordinate axis with rows representing \highlightcolor{(a)} Homogeneous, \highlightcolor{(b)} No Vernix, \highlightcolor{(c)} Back 3mm, \highlightcolor{(d)} Chest 3mm and \highlightcolor{(e)} Head 3mm volume conductor models. \protect\myfilledcircle{sensor1} indicates position of the fetal source and tissue types are color coded as follows: \protect\mycbox{fetus} = fetus, \protect\mycbox{vernix} = vernix caseosa, \protect\mycbox{amniotic} = amniotic fluid, \protect\mycbox{abdomen} = maternal body. Note: The color scale is asymmetric around 0 V/Am, indicating maximum negative potentials are approximately twice as large as maximum positive potentials.}
	\label{models}
\end{figure*}

To visualize the effects of volume conductor asymmetry in the three selected models, surface potentials produced by dipoles along each coordinate axis for the homogeneous, No Vernix, Back 3mm, Chest 3mm and Head 3mm models are shown in \figref{models}. As indicated by the Region Assessment results, these visualizations show: 1) large change in distribution patterns and greatly reduced magnitude for x axis dipoles in the Back 3mm model, 2) reduced magnitude for x axis dipoles in the Chest 3mm model, 3) large change in distribution patterns for z axis dipoles in all non-homogeneous models and 4) reduced magnitude for y axis dipoles in all non-homogeneous models.

\begin{figure*}
	\centering
	\includegraphics[width=0.99\textwidth]{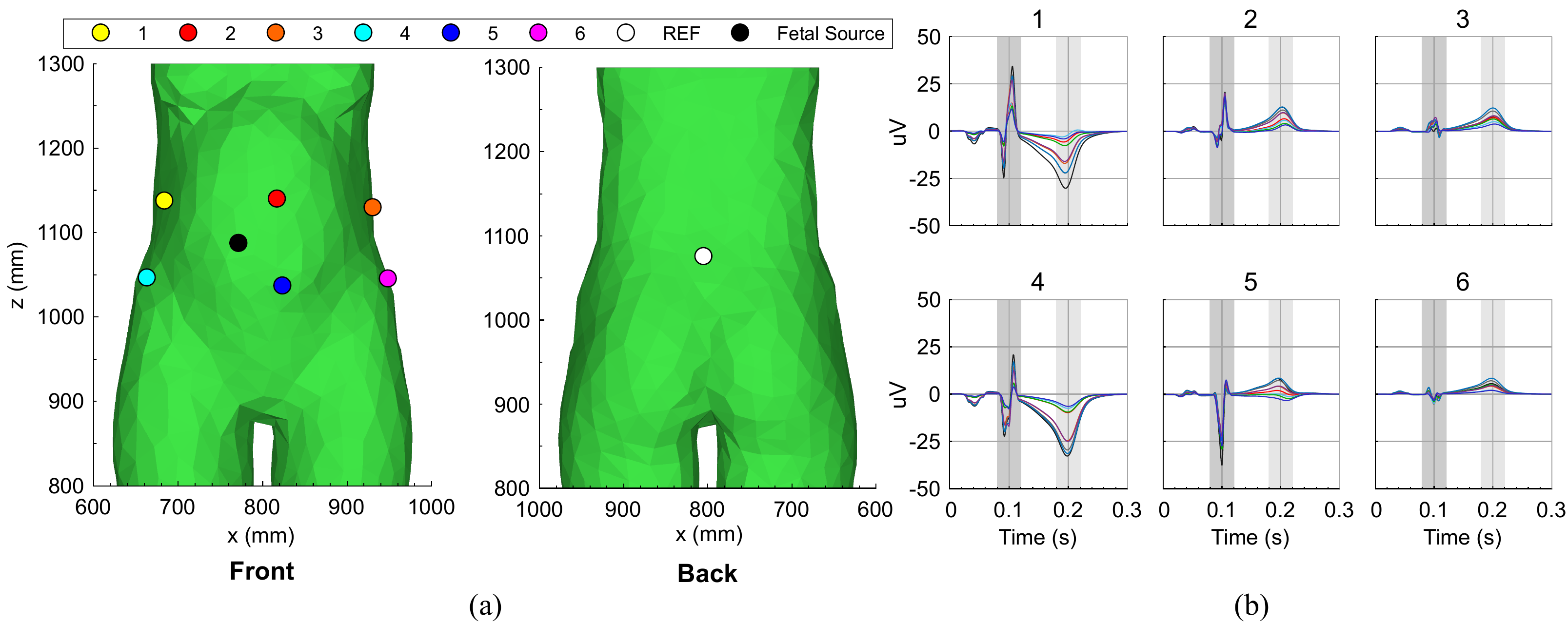}
	\caption{Simulated NI-FECG waveforms for the homogeneous and asymmetric volume conductor models \highlightcolor{(a)} shows maternal body model with 6 sensor positions (\protect\myfilledcircle{sensor7},\protect\myfilledcircle{sensor5},\protect\myfilledcircle{sensor3},\protect\myfilledcircle{sensor4},\protect\myfilledcircle{sensor2},\protect\myfilledcircle{sensor6}), reference node (\protect\myfilledcircle{sensor8}) and fetal source position (\protect\myfilledcircle{sensor1}) \highlightcolor{(b)} shows potentials observed at each sensor position with respect to the reference node for the homogeneous (\protect\mysolidline{homogeneous}) and asymmetric volume conductor models: No Vernix (\protect\mysolidline{model1}), Back 3mm (\protect\mysolidline{model4}), \mbox{Chest 3mm (\protect\mysolidline{model3})}, Head 3mm (\protect\mysolidline{model2}), BackChest 3mm (\protect\mysolidline{model7}), BackHead 3mm (\protect\mysolidline{model6}),  ChestHead 3mm (\protect\mysolidline{model5}), BackChestHead 3mm (\protect\mysolidline{model8}). \protect\mycbox{qrszone} indicates the QRS complex detection zone and  \protect\mycbox{tzone} indicates the T wave detection zone.}
	\label{waveform_results}
\end{figure*}

Following these observations, we simulate NI-FECG waveforms in volume conductor models comprising the following 9 scenarios: 1) homogeneous volume conductor model as previously defined and 2-9) finite element volume conductor models with binary combinations of vernix caseosa at 3mm thickness in the Back, Chest and Head regions where the presence/absence of vernix caseosa in each region is indicated by the model name (e.g. BackChest 3mm model contains vernix caseosa at 3mm thickness in the Back and Chest region, but not the Head region). The 6 sensor positions are equally distributed over the abdominal surface above and below the fetal source and aligned to the closest available nodes in the model discretization. Each sensor is measured in respect to a reference node at the center of the maternal back as per the default \textit{fecgsyn} sensor configuration. \figref{waveform_results} presents simulated NI-FECG waveforms for the 9 volume conductor models as described with $e_{TQRS}$ and $RDM^*$ reported at each sensor position in \tableref{signal_morphology}. To enable simple visual comparison between models, each signal's isoelectric line in \figref{waveform_results} is aligned to $0V$ and the fetal vectorcardiogram amplitude is linearly scaled by a factor of $10^{-5}$ to a peak source strength of approximately $18\mu A$.

As shown in \tableref{signal_morphology}, a maximum $e_{TQRS}$ of -0.77 is observed in the BackChest 3mm model for sensor 5, indicating 77\% relative error in the observed T/QRS ratio. The next three highest $e_{TQRS}$ values are present in the BackChestHead 3mm (-0.73), BackHead 3mm (-0.67) and Back 3mm (-0.66) models indicating that volume conductor asymmetry in the Back region has a large effect on the observed T/QRS ratio. Furthermore, in all models with vernix caseosa in the Back region, $e_{TQRS}$ is negative at all sensor positions except for

\noindent sensor 6 in the BackHead 3mm model. The No Vernix model also shows large variation in $e_{TQRS}$ with a minimum of \mbox{-0.26} in sensor 3 and maximum of 0.37 in sensor 5 indicating that volume conductor effects have a large influence on signal morphology even in the absence of vernix caseosa.

$RDM^*$ compared to a homogeneous model also demonstrates considerable variation across all models with a maximum $RDM^*$ of 0.80 observed in sensor 1 for the BackChest 3mm model followed by the BackChestHead 3mm (0.74), BackHead 3mm (0.61) and Back 3mm (0.49) models with all other models having at least one sensor with an $RDM^*$ of 0.25 or greater.
\newpage

\section{Discussion}

In this study, we aimed to characterise the effects of volume conductor asymmetry on NI-FECG extraction. To achieve this, a process was developed to compute lead field matrices and NI-FECG waveforms in a set of asymmetric volume conductor models including spatially variable distributions of vernix caseosa. Using this process, we classified the three fetal body regions with the greatest impact in terms of changes to $RDM^*$ and $lnMAG$ and assessed the impact of varying vernix caseosa thickness in each region. Based on these results, surface potentials produced by dipoles along each coordinate axis and NI-FECG waveforms at 6 sensor positions on the maternal abdomen were assessed in a range of asymmetric volume conductor models demonstrating significant changes to T/QRS ratio error and overall signal morphology compared to a homogeneous volume conductor model.
\vspace*{0.2cm}

\subsection{Region Assessment}
We observed that volume conductor asymmetry in the Back region leads to the greatest changes in terms of both $RDM^*$ and $lnMAG$, which can be attributed to its large coverage area and close proximity to the fetal source. Additionally, varying the thickness of vernix caseosa isolated in each region from 1mm to 3mm did not greatly affect $RDM^*$ or $lnMAG$ values along all coordinate axes. This can be attributed to the fact that as each region in isolation does not fully enclose the fetus, the path of least resistance to the maternal abdomen is only slightly modified by an increase in vernix caseosa thickness. From these studies, the three models of greatest impact were identified as the Back 3mm, Chest 3mm and Head 3mm models.

\subsection{Signal Morphology}
Visualisation of surface potentials produced by dipoles along each coordinate axis for the identified three models as shown in \figref{models} demonstrated large changes in potential distribution and magnitude compared to a homogeneous volume conductor model. Specifically, we observed that volume conductor asymmetry in the Back region resulted in greatly reduced magnitude for x axis dipoles, typically leading to reduced T wave amplitude as shown in \figref{waveform_results}. Following this observation, we note that difficulties in detecting T-waves via NI-FECG have been reported throughout the literature as summarised in \cite{Wacker-Gussmann_Fetalcardiactime_2017}. While other causes may likely contribute to this phenomenon, the high level of vernix caseosa asymmetry reported in the back region by Akiba \cite{Akiba_Studiesbiologicalactions_1955}, Archana \cite{Archana_ClinicalStudySurface_2008} and Visscher et al. \cite{Visscher_VernixCaseosaNeonatal_2005} presents a novel explanation for this effect. In addition, signal morphology and T/QRS ratio error were significantly altered in the No Vernix model, indicating that volume conductor effects play an important role even when vernix caseosa is not present. 

\begin{table}
	\renewcommand{\arraystretch}{1.2}
	\caption{Signal Morphology Results Indicating $e_{TQRS}$ and RDM\textsuperscript{*} At Each Sensor Position Per Model. Worst Case Per Row in \newline Bold.}
	\label{signal_morphology}
	\centering
	\begin{tabular}{C{0.118\textwidth} C{0.045\textwidth} C{0.023\textwidth} C{0.023\textwidth} C{0.023\textwidth} C{0.023\textwidth} C{0.023\textwidth} C{0.023\textwidth}}
		\hline
		\hline
		\bfseries Model & \bfseries Metric & \multicolumn{6}{c}{\bfseries Sensor Position} \\
		& & 1 & 2 & 3 & 4 & 5 & 6 \\
		\hline
		\noalign{\vskip 0.5ex} 
		\multirow{2}{*}{No Vernix} & $e_{TQRS}$ & \mbox{\hspace*{-0.093cm}-0.12} & \mbox{0.12} & \mbox{\hspace*{-0.093cm}-0.26} & \mbox{0.08} & \mbox{\textbf{0.37}} & 0.07 \\
		& RDM\textsuperscript{*} & 0.10 & \textbf{0.25} & 0.18 & 0.03 & 0.18 & 0.17 \\
		\noalign{\vskip 0.5ex}
		\hline
		\noalign{\vskip 0.5ex}
		\multirow{2}{*}{Back 3mm}& $e_{TQRS}$ & \mbox{\hspace*{-0.093cm}-0.49} & \mbox{\hspace*{-0.093cm}-0.45} & \mbox{\hspace*{-0.093cm}-0.35} & \mbox{\hspace*{-0.093cm}-0.01} & \mbox{\hspace*{-0.093cm}\textbf{-0.66}} & \mbox{\hspace*{-0.093cm}-0.20} \\
		& RDM\textsuperscript{*} & \textbf{0.49} & 0.45 & 0.27 & 0.14 & 0.36 & 0.13 \\
		\noalign{\vskip 0.5ex}
		\hline
		\noalign{\vskip 0.5ex}
		\multirow{2}{*}{Chest 3mm} & $e_{TQRS}$ & \mbox{\hspace*{-0.093cm}-0.27} & \mbox{\hspace*{-0.093cm}-0.08} & \mbox{\hspace*{-0.093cm}\textbf{-0.33}} & \mbox{0.06} & \mbox{\hspace*{-0.093cm}-0.23} & \mbox{\hspace*{-0.093cm}-0.23}\\
		& RDM\textsuperscript{*} & 0.22 & \textbf{0.26} & 0.24 & 0.05 & 0.14 & 0.16 \\
		\noalign{\vskip 0.5ex}
		\hline
		\noalign{\vskip 0.5ex}
		\multirow{2}{*}{Head 3mm} & $e_{TQRS}$ & \mbox{\hspace*{-0.093cm}-0.11} & 0.30 & \mbox{\hspace*{-0.093cm}-0.27} & \mbox{0.12} & \mbox{\textbf{0.51}} & 0.44 \\
		& RDM\textsuperscript{*} & 0.08 & 0.23 & 0.18 & 0.07 & \textbf{0.25} & 0.23 \\
		\noalign{\vskip 0.5ex}
		\hline
		\noalign{\vskip 0.5ex}
		\multirow{2}{*}{BackChest 3mm}& $e_{TQRS}$ & \mbox{\hspace*{-0.093cm}-0.71} & \mbox{\hspace*{-0.093cm}-0.52} & \mbox{\hspace*{-0.093cm}-0.46} & \mbox{\hspace*{-0.093cm}-0.14} & \mbox{\hspace*{-0.093cm}\textbf{-0.77}} & \mbox{\hspace*{-0.093cm}-0.55} \\
		& RDM\textsuperscript{*} & \textbf{0.80} & 0.52 & 0.39 & 0.20 & 0.56 & 0.45\\
		\noalign{\vskip 0.5ex}
		\hline
		\noalign{\vskip 0.5ex}
		\multirow{2}{*}{BackHead 3mm} & $e_{TQRS}$ & \mbox{\hspace*{-0.093cm}-0.30} & \mbox{\hspace*{-0.093cm}-0.65} & \mbox{\hspace*{-0.093cm}-0.25} & \mbox{\hspace*{-0.093cm}-0.04} & \mbox{\hspace*{-0.093cm}\textbf{-0.67}} & \mbox{0.04} \\
		& RDM\textsuperscript{*} & 0.23 & \textbf{0.61} & 0.20 & 0.13 & 0.59 & 0.07 \\
		\noalign{\vskip 0.5ex}
		\hline
		\noalign{\vskip 0.5ex}
		\multirow{2}{*}{ChestHead 3mm} & $e_{TQRS}$ & \mbox{\hspace*{-0.093cm}-0.27} & \mbox{\hspace*{-0.093cm}-0.16} & \mbox{\hspace*{-0.093cm}-0.44} & \mbox{0.11} & \mbox{\hspace*{-0.093cm}-0.26} & \mbox{\textbf{0.48}} \\
		& RDM\textsuperscript{*} & 0.21 & 0.29 & \textbf{0.36} & 0.15 & 0.24 & 0.15 \\
		\noalign{\vskip 0.5ex}
		\hline
		\noalign{\vskip 0.5ex}
		\multirow{2}{*}{\hspace*{-0.12cm}BackChestHead 3mm} & $e_{TQRS}$ & \mbox{\hspace*{-0.093cm}-0.55} & \mbox{\hspace*{-0.093cm}\textbf{-0.73}} & \mbox{\hspace*{-0.093cm}-0.61} & \mbox{\hspace*{-0.093cm}-0.26} & \mbox{\hspace*{-0.093cm}-0.52} & \mbox{\hspace*{-0.093cm}-0.39} \\
		& RDM\textsuperscript{*} & 0.51 & \textbf{0.74}$^\dagger$ & 0.51 & 0.29 & \textbf{0.74}$^\dagger$ & 0.32 \\
		\noalign{\vskip 0.5ex}
		\hline
		\hline
		\multicolumn{6}{l}{$\dagger$ indicates multiple columns with equal value.}
	\end{tabular}
\end{table}

Based on these observations, we conclude NI-FECG algorithm benchmarks utilizing a homogeneous volume conductor model do not indicate representative accuracy of T/QRS ratio extraction in real data and volume conductor effects should be considered in future benchmarks using simulated data. As such, without knowledge of the influence of volume conductor effects, clinical evaluation of the T/QRS ratio derived via NI-FECG should be avoided.

\begin{table*}
	\renewcommand{\arraystretch}{1.2}
	\caption{Model Refinement Results Showing $RDM^*$ and $lnMAG$ Along Each Coordinate Axis Compared To Reference Solution, Tetrahedralization Computation Time (TetGen), Lead Field Matrix Computation Time (SimBio) and Number of Nodes and Tetrahedra Per Model. Worst Case Per Column In Bold For Single Region and Multiple Regions Separately.}
	\label{model_refinement_results}
	\centering
	\begin{tabular}{C{0.01\textwidth} C{0.125\textwidth} C{0.042\textwidth} C{0.042\textwidth} C{0.042\textwidth} C{0.042\textwidth} C{0.042\textwidth} C{0.042\textwidth} C{0.07\textwidth} C{0.07\textwidth} C{0.08\textwidth} C{0.11\textwidth}} 
		\hline
		\hline
		\bfseries Type & \bfseries Model & \multicolumn{3}{c}{\bfseries RDM\textsuperscript{*}} & \multicolumn{3}{c}{\bfseries lnMAG} & \bfseries TetGen (s) & \bfseries SimBio (s) & \bfseries \# of Nodes & \bfseries \# of Tetrahedra \\
		& & x & y & z & x & y & z & & Per Sensor & ($\times10^6$) &($\times10^6$) \\
		\hline
		\noalign{\vskip 1ex}    
		\multirow{15}{*}{\rotatebox{90}{\textit{Single Region}}} & No Vernix & 0.0036 & 0.0066 & 0.0061 & 0.0011 & \mbox{\hspace*{-0.093cm}-0.0008} & 0.0008 & 91.65 & 119.86 & 0.7221 & 4.3394 \\
		& Arms 1mm & 0.0063 & 0.0108 & 0.0037 & 0.0000 & \mbox{\hspace*{-0.093cm}-0.0044} & \mbox{\hspace*{-0.093cm}-0.0024} & 113.67 & 132.46 & 0.7968 & 4.7945\\
		& Arms 3mm & 0.0029 & 0.0069 & 0.0052 & \mbox{\hspace*{-0.093cm}-0.0006} & \mbox{\hspace*{-0.093cm}-0.0038} & \mbox{\hspace*{-0.093cm}-0.0031} & 111.02 & 144.86 & 0.8094 & 4.8716\\
		& Back 1mm & 0.0088 & 0.0070 & 0.0059 & 0.0004 & \mbox{\hspace*{-0.093cm}-0.0026} & 0.0016 & \textbf{115.87} & 135.40 & 0.8103 & 4.8764\\
		& Back 3mm & 0.0036 & 0.0048 & 0.0034 & 0.0006 & \mbox{\hspace*{-0.093cm}-0.0001} & \mbox{\hspace*{-0.093cm}-0.0003} & 109.47 & 139.09 & 0.8280 & 4.9858\\
		& Buttocks 1mm & 0.0062 & 0.0063 & 0.0026 & 0.0009 & 0.0029 & 0.0009 & 95.77 & 119.38 & 0.7341 & 4.4133\\
		& Buttocks 3mm & 0.0058 & 0.0067 & 0.0036 & 0.0018 & 0.0038 & 0.0016 & 96.60 & 115.88 & 0.7367 & 4.4290\\
		& Chest 1mm & 0.0068 & 0.0040 & 0.0053 & \textbf{0.0043} & 0.0024 & 0.0033 & 104.68 & 124.57 & 0.7363  & 4.4264\\
		& Chest 3mm & \textbf{0.0133} & 0.0087 & 0.0039 & \mbox{\hspace*{-0.093cm}-0.0029} & \mbox{\hspace*{-0.093cm}-0.0028} & \mbox{\hspace*{-0.093cm}-0.0035} & 100.37 & 116.08 & 0.7384 & 4.4387\\
		& Head 1mm & 0.0053 & 0.0096 & 0.0042 & \mbox{\hspace*{-0.093cm}-0.0002} & \mbox{\hspace*{-0.093cm}-0.0003} & 0.0008 & 104.89 & 132.79 &  0.8219 & 4.9478\\
		& Head 3mm & 0.0071 & 0.0110 & 0.0032 & \mbox{\hspace*{-0.093cm}-0.0018} & 0.0023 & \mbox{\hspace*{-0.093cm}-0.0005} & 109.39 & \textbf{156.95} & \textbf{0.8449} & \textbf{5.0884}\\
		& Inguinal 1mm & 0.0084 & 0.0115 & \textbf{0.0094} & 0.0022 & \textbf{0.0080} & \textbf{0.0073} & 108.30 & 110.73 & 0.7275 & 4.3716\\
		& Inguinal 3mm & 0.0071 & \textbf{0.0128} & 0.0052 & 0.0009 & \mbox{\hspace*{-0.093cm}-0.0048} & \mbox{\hspace*{-0.093cm}-0.0002} & 96.29 & 116.43 & 0.7291 & 4.3824\\
		& Legs 1mm & 0.0069 & 0.0020 & 0.0035 & \mbox{\hspace*{-0.093cm}-0.0017} & 0.0015 & \mbox{\hspace*{-0.093cm}-0.0026} & 113.86 & 135.69 & 0.8056 & 4.8471\\
		& Legs 3mm & 0.0041 & 0.0077 & 0.0043 & 0.0004 & \mbox{\hspace*{-0.093cm}-0.0034} & 0.0008 & 108.24 & 151.92 & 0.8192 & 4.9303\\ \noalign{\vskip 1ex} 
		\hline
		\noalign{\vskip 1ex} 
		\multirow{4}{*}{\rotatebox{90}{\textit{Mult. Regions}}} & \mbox{BackChest 3mm} & 0.0164 & 0.0038 & \textbf{0.0069} & \textbf{0.0117} & 0.0006 & 0.0021 & 113.63 & 150.34 & 0.8435 & 5.0795 \\
		& BackHead 3mm & 0.0073 & 0.0045 & 0.0068 & 0.0080 & \mbox{\hspace*{-0.093cm}-0.0029} & \mbox{\hspace*{-0.093cm}-0.0005} & 135.31 & 173.86 & 0.9521 & 5.7427\\
		& ChestHead 3mm & 0.0114 & 0.0035 & 0.0056 & \mbox{\hspace*{-0.093cm}-0.0022} & \mbox{\hspace*{-0.093cm}-0.0020} & 0.0028 & 125.54 & 159.00 & 0.8606 & 5.1845\\
		& \mbox{\hspace*{-0.093cm}BackChestHead 3mm} & \textbf{0.0204} & \textbf{0.0053} & 0.0041 & 0.0028 & \textbf{0.0034} & \textbf{0.0060} & \textbf{139.47} & \textbf{185.91} & \textbf{0.9672} & \textbf{5.8341}\\
		\noalign{\vskip 1ex} 
		\hline
		\hline
	\end{tabular}
\end{table*}

\subsection{Future Work}

In the present study, we considered a single fetal model with a fixed set of source parameters. While these parameters were carefully selected based on clinical measurements, further analysis should be conducted using a range of anatomic and vectorcardiogram models to determine the typical distribution of error within a larger pregnancy cohort. Further clinical studies are also warranted to determine the likelihood of the investigated vernix caseosa distributions across different gestational ages and validate the predictions of our model against observed abdominal potentials. Regardless of these factors, the presented results demonstrate the significant potential for estimation error in real data and confirm that volume conductor effects represent an important topic in the study of NI-FECG extraction accuracy. 

In addition to the 32 weeks GA model utilized in this work, the developed process allows for the simulation of NI-FECG waveforms using any triangular PLC, enabling the study of volume conductor effects over a range of gestational ages and fetal positions with the possibility to include additional structures in the modeling process such as the placenta, umbilical cord and internal organs. Furthermore, as the surface potentials in our process are generated according to a linear mixture, it can be utilized to simulate abdominal ECG mixtures for multiple fetuses (e.g. twins) via the inclusion of additional cardiac sources within an appropriate volume conductor model.

Finally, our process can be easily extended to the field of fetal magnetocardiography, allowing study of the effects of volume conductor asymmetry on magnetic field distribution. As our process has been developed using MATLAB and a set of open-source tools, a future development goal is to release the developed code under an open-source license to enable rapid NI-FECG benchmarks incorporating volume conductor effects within the research community. 

\section{Conclusion}
 This work demonstrates that volume conductor effects have a significant impact on the signal morphology derived via NI-FECG. Simulation studies using finite element models of the maternal-fetal anatomy revealed that volume conductor asymmetry can result in over 70\% error in the observed T/QRS ratio and significant morphological changes compared to a homogeneous volume conductor model. Future NI-FECG algorithm benchmarks using simulated data should incorporate volume conductor effects to provide better indication of NI-FECG extraction accuracy using real data.

\section*{Appendix}
Model refinement results are shown in \tableref{model_refinement_results} for the \textit{abdominal sensors} lead field matrix for all utilized finite element models. For each model, \tableref{model_refinement_results} describes $RDM^*$ and $lnMAG$ along each coordinate axis compared to a reference solution with halved maximum element volume in each compartment, tetrahedralization computation time via TetGen, lead field matrix computation time via SimBio and number of nodes and tetrahedra.

\section*{Acknowledgment}

Emerson Keenan is supported by an Australian Government Research Training Program Scholarship at the University of Melbourne.

\bibliographystyle{IEEEtran}

\bibliography{IEEEtranBSTControl,IEEEabrv,references}

\end{document}